\documentclass[nofootinbib, pra,  twocolumn,superscriptaddress]{revtex4-1}

\usepackage[T1]{fontenc}
\usepackage[latin9]{inputenc}
\usepackage{color}
\usepackage{mathrsfs}
\usepackage{amsmath}
\usepackage{amssymb}
\usepackage{graphicx}
\usepackage[unicode=true,pdfusetitle,
 bookmarks=true,bookmarksnumbered=false,bookmarksopen=false,
 breaklinks=false,pdfborder={0 0 0},pdfborderstyle={},backref=false,colorlinks=true]
 {hyperref}
\hypersetup{
 citecolor=blue,urlcolor=blue}

\makeatletter
\usepackage{bbm}

\makeatother

\begin{document}
\title{Quantum refrigerator driven by nonclassical light}
\author{Hui-Jing Cao}
\affiliation{Center for Quantum Technology Research, and Key Laboratory of Advanced
Optoelectronic Quantum Architecture and Measurements, School of Physics,
Beijing Institute of Technology, Beijing 100081, China}
\author{Fu Li}
\affiliation{Department of Electrical and Systems Engineering, Washington University,
St. Louis, Missouri 63130, USA}
\author{Sheng-Wen Li}
\email{lishengwen@bit.edu.cn}

\affiliation{Center for Quantum Technology Research, and Key Laboratory of Advanced
Optoelectronic Quantum Architecture and Measurements, School of Physics,
Beijing Institute of Technology, Beijing 100081, China}
\begin{abstract}
We study a three-level quantum refrigerator which is driven by a generic
light state, even a nonclassical one. With the help of P function
expansion of the driving light, we obtain the heat current generated
by different types of light states. It turns out all different input
light states give the same coefficient of performance for this refrigerator,
while the cooling power depend not only on the light intensity but
also the specific photon statistics of the driving light. Comparing
with the coherent light with the same intensity, the driving light
with super(sub)-Poissonian photon statistics could raise a smaller
(stronger) cooling power. We find that this is because the bunching
photons would first excite the system but then successively induce
the stimulated emission, which draws the refrigerator back to the
starting state of the cooling process and thus decreases the cooling
current generation. This mechanism provides a more delicate control
method via the high order coherence of the input light. 
\end{abstract}
\maketitle

\section{Introduction}

When a quantum heat engine runs between two reservoirs containing
specific quantum coherences, the efficiency of the heat engine could
exceed the Carnot limit between two canonical thermal baths \citep{scully_quantum_2002,scully_extracting_2003,rosnagel_nanoscale_2014,manzano_entropy_2016}.
But such exotic effects are restricted for practical applications
since quantum coherences are usually quite fragile confronting the
noises surrounded. In contrast, applying a quantum light to control
or drive a quantum refrigerator is feasible under current techniques,
promising intriguing properties. Some recent studies show that, comparing
with the normal laser light with the same intensity, using nonclassical
squeezed light could help enhance the two-photon absorption rate \citep{li_squeezed_2020},
and exceed the cooling limit in laser cooling experiments \citep{schafermeier_quantum_2016,clark_sideband_2017}. 

Therefore, here we study a quantum refrigerator which is driven by
different types of light states, especially, the nonclassical lights.
We focus on a typical quantum thermal machine composed of a three-level
system in contact with two heat baths \citep{scovil_three-level_1959}.
Applying a proper temperature difference to the two baths, a population
inversion is generated between two levels, and the system could work
as a heat engine, emitting laser light as its output work \citep{boukobza_thermodynamic_2006,boukobza_three-level_2007,li_quantum_2017,kosloff_quantum_2017}.
Reversely, when a driving light is input to the three-level system,
it works as a refrigerator {[}Fig.\,\ref{fig-3level}(a){]}, moving
the heat from the cold bath to the hot one \citep{geusic_quantum_1967}.
It is also worth noting that the transition structure of this three-level
refrigerator is analogous to many other physical systems, such as
the sideband cooling system {[}Fig.\,\ref{fig-3level}(c){]} \citep{tian_ground_2009},
and photovoltaic systems \citep{rutten_reaching_2009,wang_optimal_2014}.

When the driving light shining on the refrigerator is a generic quantum
state, it is no longer enough to treat the driving light simply as
a planar wave, which is a quasi-classical description in literatures.
To study the interaction with a nonclassical driving light, notice
that, with the help of the P function representation, a generic light
state can be regarded as the combination of many coherent states $|\alpha\rangle$
with $P(\alpha,\alpha^{*})$ as the ``quasi-probability'', while
the coherent states $|\alpha\rangle$ are the quantum correspondences
for the classical planar waves \citep{orszag_quantum_2000,vogel_quantum_2006,scully_quantum_1997,agarwal_quantum_2012}.
Therefore, the full system dynamics can be obtained as the P function
average of many evolution ``branches'', and each evolution branch
can be obtained from the above quasi-classical approach, treating
the driving light as a planar wave \citep{ritsch_atomic_1988,ritsch_systems_1988,gardiner_driving_1994,yao_enhancing_2020}.

Based on this approach, we obtain the cooling power of this quantum
refrigerator for different input light states. It turns out the coefficient
of performance (COP) always remains as $e=\omega_{\mathrm{c}}/(\omega_{\mathrm{h}}-\omega_{\mathrm{c}})\le T_{\mathrm{c}}/(T_{\mathrm{h}}-T_{\mathrm{c}})$,
whose upper bound is just the Carnot limit for refrigerators. But
the cooling powers generated by different driving lights depend on
the specific photon statistics. Comparing with the coherent light
with the same intensity, the driving light with super(sub)-Poissonian
photon statistics would raise a smaller (stronger) cooling power. 

We find that, this is because bunching photons would block the cooling
current generation due to the stimulated emission they bring in. When
a pair of bunching photons income together, the first photon would
excite the refrigerator system up, but the second photon successively
followed would induce the stimulated emission, drawing the system
back to the previous state, and that blocks the generation of the
cooling current flowing to the hot bath. Therefore, comparing with
the coherent light where the photons income randomly both in bunches
and individually, the bunching (antibunching) light could generate
a smaller (stronger) cooling power under the same light intensity.
Clearly, the similar mechanism could also take effect in many other
systems undergoing light driving. 

As a comparison, we also consider the situation that the whole multimode
light field is in the thermal equilibrium state with a temperature
$T_{\text{\textsc{e}}}$ \citep{skrzypczyk_smallest_2011,chen_quantum_2012,levy_quantum_2012,wang_four-level_2015}.
It turns out the thermal photon number must be larger than a certain
threshold so as to make sure the system works as a refrigerator. Again
that indicates the working status of the system is determined not
only by the light intensity but also the specific quantum state of
the light field.

The paper is arranged as follows. In Sec.\,II we show the basic properties
of the three-level system quantum refrigerator under a coherent driving
light. In Sec.\,III we discuss the situation that the driving light
carries a generic photon statistics. In Sec.\,IV we show that the
driving light with antibunching statistics could enhance the cooling
power. In Sec.\,V we consider the situation that the whole multimode
EM field is in a thermal equilibrium state. The summary is drawn in
Sec.\,VI, and some detailed derivations are placed in the Appendices.

\section{The three-level quantum refrigerator}

\begin{figure}
\includegraphics[width=1\columnwidth]{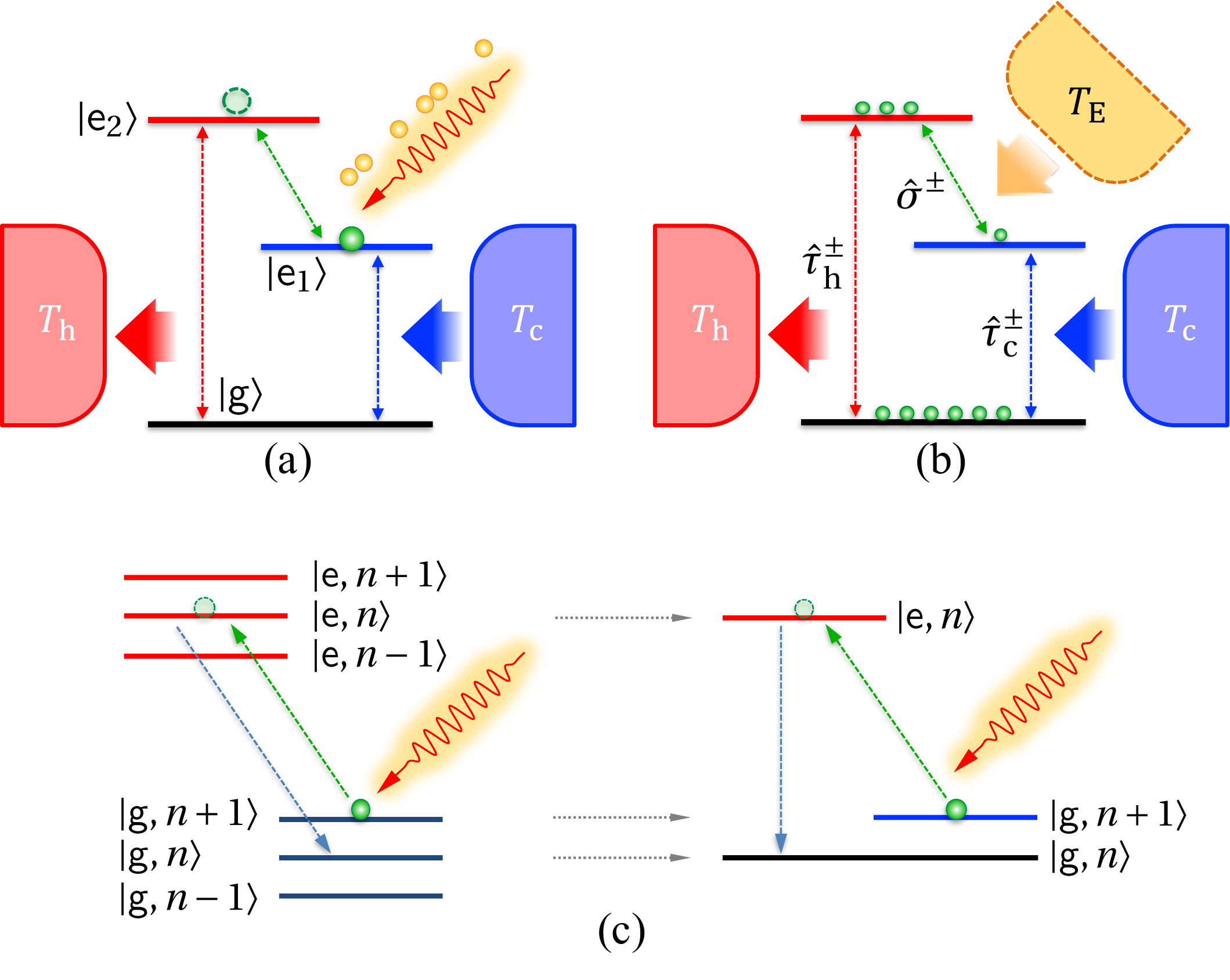}\caption{Demonstrations for the the quantum refrigerator. (a) The transition
$|\mathsf{e}_{1}\rangle\leftrightarrow|\mathsf{e}_{2}\rangle$ is
driven by an idealistic single mode light, which may carry nonclassical
photon statistics. (b) The the whole EM field is in the canonical
thermal state with temperature $T_{\text{\textsc{e}}}$. (c) An analogue
between this three-level quantum refrigerator and the energy level
structure in the sideband cooling systems, where the oscillation motion
(of an ion or mechanical oscillator) is cooled down with the help
of a two-level system ($|\mathsf{e}\rangle$, $|\mathsf{g}\rangle$)
coupled with it, and $|n\rangle$ indicates the phonon states.}
\label{fig-3level}
\end{figure}

The basic setup of the three-level quantum refrigerator is shown in
Fig.\,\ref{fig-3level}(a), which is described by the Hamiltonian
$\hat{H}_{\textrm{s}}=\text{\textsc{e}}_{1}|\mathsf{e}_{1}\rangle\langle\mathsf{e}_{1}|+\text{\textsc{e}}_{2}|\mathsf{e}_{2}\rangle\langle\mathsf{e}_{2}|$
(the ground state energy is set as 0). The transition pathways $|\mathsf{e}_{1}\rangle\leftrightarrow|\mathsf{g}\rangle$
and $|\mathsf{e}_{2}\rangle\leftrightarrow|\mathsf{g}\rangle$ are
coupled with two independent bosonic heat baths ($\hat{H}_{\text{\textsc{b}},i}=\sum_{k}\omega_{i,k}\hat{b}_{i,k}^{\dagger}\hat{b}_{i,k}$,
for $i=\mathrm{h},\mathrm{c}$), and their interaction Hamiltonians
are $\hat{H}_{\text{\textsc{sb}},i}=\hat{\tau}_{i}^{+}\cdot\sum_{k}g_{i,k}\hat{b}_{i,k}+\mathbf{h.c.}$,
with $\hat{\tau}_{\mathrm{h(c)}}^{+}:=|\mathsf{e}_{2(1)}\rangle\langle\mathsf{g}|=[\hat{\tau}_{\mathrm{h(c)}}^{-}]^{\dagger}$
as the transition operator. We consider the two heat baths stay in
the thermal equilibrium states with the temperatures $T_{\mathrm{h}}>T_{\mathrm{c}}$.
It is worth noting that indeed the basic transition structure of the
sideband cooling system is quite similar as this three-level model
{[}Fig.\,\ref{fig-3level}(c){]}.

Here we show that, when using a light beam to drive the transition
 $|\mathsf{e}_{1}\rangle\leftrightarrow|\mathsf{e}_{2}\rangle$, such
a three level system could work as a quantum refrigerator, namely,
the net energy flux would flow from the cold bath to the hot one \citep{geusic_quantum_1967,wang_four-level_2015}.

Generally, the driving light is modeled as a classical planar wave
with a single frequency mode, and its interaction with the three level
system is described by \citep{orszag_quantum_2000,vogel_quantum_2006,scully_quantum_1997,agarwal_quantum_2012}
\begin{align}
\hat{V}(t) & =-\hat{\boldsymbol{d}}\cdot\vec{E}_{\mathrm{d}}\,\sin(\omega_{\mathrm{d}}t-\phi_{\mathrm{d}}).\label{eq:1}
\end{align}
where $\omega_{\mathrm{d}}$ is the driving frequency, $\hat{\boldsymbol{d}}:=\vec{\wp}\,\hat{\sigma}^{-}+\mathbf{h.c.}$
is the dipole operator of the three level system, with $\vec{\wp}:=\langle\mathsf{e}_{1}|\hat{\boldsymbol{d}}|\mathsf{e}_{2}\rangle$
as the transition dipole moment, and $\hat{\sigma}^{-}:=|\mathsf{e}_{1}\rangle\langle\mathsf{e}_{2}|:=(\hat{\sigma}^{+})^{\dagger}$.

Under the Born-Markovian-rotating-wave approximation \citep{breuer_theory_2002},
the dynamics of the system can be described by the following master
equation\footnote{Hereafter, $\tilde{o}$ indicates the operator
in the interaction picture of $\hat{H}_{\text{\textsc{s}}}$.}
\begin{align}
\partial_{t}\tilde{\rho} & =\frac{i}{\hbar}[\tilde{\rho},\tilde{V}(t)]+\mathcal{L}_{\text{\textsc{e}}}[\tilde{\rho}]+\mathcal{L}_{\mathrm{c}}[\tilde{\rho}]+\mathcal{L}_{\mathrm{h}}[\tilde{\rho}],\label{eq:ME}\\
\tilde{V} & =i\mathcal{E}\hat{\sigma}^{+}e^{i\Delta t}-i\mathcal{E}^{*}\hat{\sigma}^{-}e^{-i\Delta t},\nonumber 
\end{align}
where $\Delta:=\Omega-\omega_{\mathrm{d}}$ is the detuning between
the driving light and the transition frequency $\hbar\Omega:=\text{\textsc{e}}_{2}-\text{\textsc{e}}_{1}$
, $\mathcal{E}:=e^{i\phi_{\mathrm{d}}}(\vec{\wp}\cdot\vec{E}_{\mathrm{d}})/2$
is the driving strength, and the dissipations terms are 
\begin{align}
\mathcal{L}_{\text{\textsc{e}}}[\tilde{\rho}] & =\kappa\big(\hat{\sigma}^{-}\tilde{\rho}\hat{\sigma}^{+}-\frac{1}{2}\{\tilde{\rho},\,\hat{\sigma}^{+}\hat{\sigma}^{-}\}_{+}\big),\nonumber \\
\mathcal{L}_{i}[\tilde{\rho}] & =\gamma_{i}\bar{\mathsf{n}}_{i}\big(\hat{\tau}_{i}^{+}\tilde{\rho}\hat{\tau}_{i}^{-}-\frac{1}{2}\{\tilde{\rho},\,\hat{\tau}_{i}^{-}\hat{\tau}_{i}^{+}\}_{+}\big)\nonumber \\
 & +\gamma_{i}(\bar{\mathsf{n}}_{i}+1)\big(\hat{\tau}_{i}^{-}\tilde{\rho}\hat{\tau}_{i}^{+}-\frac{1}{2}\{\tilde{\rho},\,\hat{\tau}_{i}^{+}\hat{\tau}_{i}^{-}\}_{+}\big),\label{eq:Lindblad}
\end{align}
for $i=\mathrm{h},\mathrm{c}$. Here $\mathcal{L}_{\mathrm{h}(\mathrm{c})}[\tilde{\rho}]$
gives the energy exchange with the hot (cold) bath, with $\bar{\mathsf{n}}_{i}:=[\exp(\hbar\omega_{i}/k_{\text{\textsc{b}}}T_{i})-1]^{-1}$
from the Planck function (with $\hbar\omega_{\mathrm{h(c)}}\equiv\text{\textsc{e}}_{1(2)}$),
and $\mathcal{L}_{\text{\textsc{e}}}[\tilde{\rho}_{\textrm{s}}]$
gives the spontaneous emission to the EM field.

From the master equation (\ref{eq:ME}), the changing rate of the
system energy gives 
\begin{align}
\partial_{t}\langle\hat{H}_{\text{\textsc{s}}}\rangle= & \mathrm{tr}\big\{\frac{i}{\hbar}[\tilde{\rho},\tilde{V}(t)]\hat{H}_{\text{\textsc{s}}}+\mathcal{L}_{\text{\textsc{e}}}[\tilde{\rho}]\hat{H}_{\text{\textsc{s}}}\big\}+\mathrm{tr}\big\{\mathcal{L}_{\mathrm{c}}[\tilde{\rho}]\hat{H}_{\text{\textsc{s}}}\big\}\nonumber \\
+ & \mathrm{tr}\big\{\mathcal{L}_{\mathrm{h}}[\tilde{\rho}]\hat{H}_{\text{\textsc{s}}}\big\}:=\mathcal{Q}_{\text{\textsc{e}}}+\mathcal{Q}_{\mathrm{c}}+\mathcal{Q}_{\mathrm{h}},\label{eq:E-Q}
\end{align}
where $\mathcal{Q}_{\text{\textsc{e}}}$, $\mathcal{Q}_{\mathrm{c}}$,
$\mathcal{Q}_{\mathrm{h}}$ are the energy flux flowing to the system
from the EM field, cold and hot baths respectively. In the steady
state $\partial_{t}\langle\hat{H}_{\text{\textsc{s}}}\rangle=0$,
the above energy flows can be obtained by solving master equation
(\ref{eq:ME}) (Appendix \ref{sec:The-power-of}). Here we consider
the situation that the spontaneous emission rate $\kappa$ is negligible
comparing with the coupling strengths with the two heat bath ($\gamma_{\mathrm{h,c}}\gg\kappa\rightarrow0$),
and that gives the heat flows as (let $\gamma_{\mathrm{h}}=\gamma_{\mathrm{c}}\equiv\gamma$)
\begin{align}
\mathcal{Q}_{\mathrm{h}} & =-\hbar\omega_{\mathrm{h}}J,\quad\mathcal{Q}_{\mathrm{c}}=\hbar\omega_{\mathrm{c}}J,\quad\mathcal{Q}_{\text{\textsc{e}}}=\hbar\Omega\,J,\nonumber \\
J & =\frac{|\vec{\wp}\cdot\vec{E}_{\mathrm{d}}|^{2}\,(\bar{\mathsf{n}}_{\mathrm{c}}-\bar{\mathsf{n}}_{\mathrm{h}})}{\Gamma_{1}\mathcal{N}+4\Delta^{2}/\Gamma_{1}+|\vec{\wp}\cdot\vec{E}_{\mathrm{d}}|^{2}\,\Gamma_{2}/\gamma^{2}}\label{eq:Q-hc}\\
 & \equiv\frac{\mathscr{A}|\vec{E}_{\mathrm{d}}|^{2}}{\mathscr{B}+\mathscr{C}|\vec{E}_{\mathrm{d}}|^{2}},\nonumber 
\end{align}
where $J$ is the population flux, $\mathscr{A}\equiv\wp_{\mathrm{d}}^{2}(\bar{\mathsf{n}}_{\mathrm{c}}-\bar{\mathsf{n}}_{\mathrm{h}})$,
$\mathscr{B}\equiv\Gamma_{1}\mathcal{N}+4\Delta^{2}/\Gamma_{1}$,
$\mathscr{C}\equiv\wp_{\mathrm{d}}^{2}\Gamma_{2}/\gamma^{2}$ are
the abbreviated coefficients (denoting $\wp_{\mathrm{d}}:=|\vec{\wp}\cdot\hat{\mathrm{e}}_{\mathrm{d}}|$,
with $\hat{\mathrm{e}}_{\mathrm{d}}$ as the unit direction of $\vec{E}_{\mathrm{d}}$),
and 
\begin{align}
\Gamma_{1} & =\gamma(2+\bar{\mathsf{n}}_{\mathrm{c}}+\bar{\mathsf{n}}_{\mathrm{h}}),\quad\Gamma_{2}=\gamma(2+3\bar{\mathsf{n}}_{\mathrm{c}}+3\bar{\mathsf{n}}_{\mathrm{h}}),\nonumber \\
\mathcal{N} & =1+2\bar{\mathsf{n}}_{\mathrm{h}}+2\bar{\mathsf{n}}_{\mathrm{c}}+3\bar{\mathsf{n}}_{\mathrm{c}}\bar{\mathsf{n}}_{\mathrm{h}}.\label{eq:Gamma}
\end{align}

Notice that, as long as $\bar{\mathsf{n}}_{\mathrm{c}}-\bar{\mathsf{n}}_{\mathrm{h}}\ge0$
with $\mathcal{E}\neq0$, we have $\mathcal{Q}_{\mathrm{c}}\ge0$
and $\mathcal{Q}_{\mathrm{h}}\leq0$, which means the heat is flowing
across the system from the cold bath to the hot one. Namely, the incoming
light is driving the system to work as a refrigerator, and the above
inequality gives the cooling condition as $\omega_{\mathrm{c}}/T_{\mathrm{c}}\le\omega_{\mathrm{h}}/T_{\mathrm{h}}$.
When the driving strength is weak, the cooling power is proportional
to the light intensity $J\simeq(\mathscr{A}/\mathscr{B})\,|\vec{E}_{\mathrm{d}}|^{2}$. 

As a result, the coefficient of performance (COP) for this three-level
refrigerator gives 
\begin{align}
e & \equiv\frac{|\mathcal{Q}_{\mathrm{c}}|}{|\mathcal{Q}_{\mathrm{h}}|-|\mathcal{Q}_{\mathrm{c}}|}=\frac{\omega_{\mathrm{c}}}{\omega_{\mathrm{h}}-\omega_{\mathrm{c}}}\le\frac{T_{\mathrm{c}}}{T_{\mathrm{h}}-T_{\mathrm{c}}}.\label{eq:cofficient}
\end{align}
Therefore, this COP is just bounded by the Carnot limit for refrigerators.
The equality holds when the energy flows {[}Eq.\,(\ref{eq:Q-hc}){]}
approach zero, which indicates the quasi-static and reversible process,
leading to the zero power. When the spontaneous emission rate $\kappa$
is a small but finite value, the upper bound of the COP would be smaller
than the Carnot limit. 

\section{The driving light with generic photon statistics }

Now we consider a more general situation that the input light driving
the refrigerator could carry different kinds of photon statistics,
but still has a quite small linewidth and can be regarded as a monochromatic
light. 

In this situation, it is no longer enough to treat the driving light
only as a classical planar wave, which cannot reflect the photon statistics
of the input light. Here we consider the EM field is fully quantized,
which is described the multimode Hamiltonian $\hat{H}_{\text{\textsc{e}}}=\sum\hbar\omega_{\mathbf{k}}\hat{a}_{\mathbf{k}\varsigma}^{\dagger}\hat{a}_{\mathbf{k}\varsigma}$.
The driving mode stays in a specific quantum state, while all the
other field modes are in the vacuum state. Generally, the quantum
state of the driving mode always can be represented as the following
P function, 
\begin{equation}
\hat{\varrho}_{\mathrm{d}}=\int d^{2}\alpha\,P(\alpha,\alpha^{*})\,|\alpha\rangle\langle\alpha|.
\end{equation}
Formally, this density state $\hat{\varrho}_{\mathrm{d}}$ of the
driving mode can be regarded as the combination of many coherent state
$|\alpha\rangle$, with $P(\alpha,\alpha^{*})$ as the quasi-probability.
For the quantized EM field, a single mode coherent state $|\alpha\rangle$
corresponds to a classical planar wave, since the electric field operator
$\hat{\mathbf{E}}(\mathbf{x},t)\equiv\hat{\mathbf{E}}_{-}+\hat{\mathbf{E}}_{+}$
gives 
\begin{align}
\hat{\mathbf{E}}_{+}(\mathbf{x},t)=(\hat{\mathbf{E}}_{-})^{\dagger} & \equiv i\sum_{\mathbf{k}\varsigma}\hat{\mathrm{e}}_{\mathbf{k}\varsigma}\sqrt{\frac{\hbar\omega_{\mathbf{k}}}{2\epsilon_{0}V}}\,\hat{a}_{\mathbf{k}\varsigma}e^{i\mathbf{k}\cdot\mathbf{x}-i\omega_{\mathbf{k}}t}\nonumber \\
\langle\alpha|\hat{\mathbf{E}}(\mathbf{x},t)|\alpha\rangle & =\vec{E}_{\alpha}\sin(\omega_{\mathrm{d}}t-\mathbf{k}_{\mathrm{d}}\cdot\mathbf{x}-\phi_{\alpha}),\label{eq:E-alpha}
\end{align}
where $\hat{\mathbf{E}}_{\pm}$ are the field operators with positive
and negative frequencies, and $\vec{E}_{\alpha}:=\hat{\mathrm{e}}_{\mathrm{d}}\,|\alpha|\sqrt{2\hbar\omega_{\mathrm{d}}/\epsilon_{0}V}$.

In this sense, the generic state $\hat{\varrho}_{\mathrm{d}}$ of
the driving light could be regarded as a ``probabilistic'' combination
of many classical planar waves. Therefore, the system dynamics also
could be obtained as the combination of many evolution ``branches'',
and in each branch the system is driven by the planar wave given by
Eq.\,(\ref{eq:E-alpha}), that is (a rigorous proof is shown in Appendix
\ref{sec:Master-equation-derivation}), 
\begin{align}
\tilde{\rho}(t) & =\int d^{2}\alpha\,P(\alpha,\alpha^{*})\,\tilde{\rho}^{(\alpha)}(t),\label{eq:ME-1}\\
\partial_{t}\tilde{\rho}^{(\alpha)} & =\frac{i}{\hbar}[\tilde{\rho}^{(\alpha)},\tilde{V}_{\alpha}(t)]+\mathcal{L}[\tilde{\rho}^{(\alpha)}],\nonumber 
\end{align}
where the master equation for $\tilde{\rho}^{(\alpha)}(t)$ has the
same form as the above Eq.\,(\ref{eq:ME}) for the quasi-classical
driving \citep{ritsch_atomic_1988,ritsch_systems_1988,gardiner_driving_1994,yao_enhancing_2020}.
Here the driving light in $\tilde{V}_{\alpha}$ should be replaced
by the planar wave in Eq.\,(\ref{eq:E-alpha}) determined by the
coherent state $|\alpha\rangle$, and the driving strength in Eq.\,(\ref{eq:ME})
now should be modified as $\mathcal{E}_{\alpha}:=e^{i\phi_{\alpha}}(\vec{\wp}\cdot\vec{E}_{\alpha})/2$
respectively. Correspondingly, the expectations of the system observables
are obtained as 
\begin{equation}
\langle\hat{o}_{\text{\textsc{s}}}(t)\rangle\equiv\mathrm{tr}[\hat{o}_{\text{\textsc{s}}}\cdot\hat{\rho}(t)]=\int d^{2}\alpha\,P(\alpha,\alpha^{*})\,\langle\hat{o}_{\text{\textsc{s}}}^{(\alpha)}(t)\rangle,\label{eq:Os-P}
\end{equation}
where $\langle\hat{o}_{\text{\textsc{s}}}^{(\alpha)}(t)\rangle:=\mathrm{tr}[\hat{o}_{\text{\textsc{s}}}\cdot\hat{\rho}^{(\alpha)}(t)]$.

Namely, the full system evolution $\langle\hat{o}_{\text{\textsc{s}}}(t)\rangle$
could be regarded as the ``probabilistic'' summation of many evolution
``branches'' $\langle\hat{o}_{\text{\textsc{s}}}^{(\alpha)}(t)\rangle$,
with $P(\alpha,\alpha^{*})$ as their probabilities, and each ``branches''
$\langle\hat{o}_{\text{\textsc{s}}}^{(\alpha)}(t)\rangle$ can be
obtained from the above master equation with the quasi-classical driving.

Based on this method, now we study the heat flows of the above quantum
refrigerator when the driving light carries generic photon statistics.
Similar as the energy-flow conservation relation (\ref{eq:E-Q}),
the three heat flows are given as $\overline{\mathcal{Q}}_{\mathrm{h}}=-\hbar\omega_{\mathrm{h}}\,\overline{J}$,
$\overline{\mathcal{Q}}_{\mathrm{c}}=\hbar\omega_{\mathrm{c}}\,\overline{J}$
and $\overline{\mathcal{Q}}_{\text{\textsc{e}}}=\hbar\Omega\,\overline{J}$,
where $\overline{J}$ is the population flux obtained from the P function
average (\ref{eq:Os-P}), namely, 
\begin{equation}
\overline{J}=\int d^{2}\alpha\,P(\alpha,\alpha^{*})\,J_{\alpha},\quad J_{\alpha}\equiv\frac{\mathscr{A}|\vec{E}_{\alpha}|^{2}}{\mathscr{B}+\mathscr{C}|\vec{E}_{\alpha}|^{2}}.\label{eq:J-P}
\end{equation}
Here $J_{\alpha}$ is the same as the above Eq.\,(\ref{eq:Q-hc}),
which indicates the steady state flux when the driving light is a
classical planar wave corresponding to the coherent state as Eq.\,(\ref{eq:E-alpha}).

As a result, in spite of the driving light state, the COP for this
refrigerator is still $e=\omega_{\mathrm{c}}/(\omega_{\mathrm{h}}-\omega_{\mathrm{c}})$
as long as $\overline{J}\ge0$, which would lead to the same cooling
condition as Eq.\,(\ref{eq:cofficient}).

\section{Enhancing the cooling power by anti-bunching photons }

As long as the driving light state $P(\alpha,\alpha^{*})$ is known,
the heat flows of the refrigerator can be obtained from the above
Eq.\,(\ref{eq:J-P}). Here we consider some typical examples of different
driving light.

We first consider the driving light is carrying the thermal statistics,
whose P function is given by $P_{\mathrm{th}}(\alpha)=(\pi\bar{n}_{\mathrm{th}})^{-1}\textrm{exp}[-|\alpha|^{2}/\bar{n}_{\mathrm{th}}]$,
with $\bar{n}_{\mathrm{th}}$ as the mean photon number. Such a distribution
is also consistent with the classical picture for the chaotic light,
which is regarded as the probabilistic combination of planar waves,
whose intensities satisfy the negative exponential distribution \citep{goodman_statistical_2000,li_photon_2020}.
In this case, from Eq.\,(\ref{eq:J-P}), the population flow in the
steady state gives 
\begin{equation}
\overline{J}_{\mathrm{th}}=\int d^{2}\alpha\,\frac{e^{-|\alpha|^{2}/\bar{n}_{\mathrm{th}}}}{\pi\bar{n}_{\mathrm{th}}}\frac{\xi_{0}^{2}|\alpha|^{2}(\bar{\mathsf{n}}_{\mathrm{c}}-\bar{\mathsf{n}}_{\mathrm{h}})}{\mathscr{B}+\xi_{0}^{2}|\alpha|^{2}\Gamma_{2}/\gamma^{2}}.\label{eq:9-Pw-th}
\end{equation}
Here $\xi_{0}:=\wp_{\mathrm{d}}\sqrt{2\hbar\omega_{\mathrm{d}}/\epsilon_{0}V}$
is the single photon coupling strength, where $V$ takes the coherence
volume of the driving light \citep{goodman_statistical_2000}. And
the intensity of the driving light (the Poynting vector) is $I_{E}\equiv2c\epsilon_{0}\langle\hat{E}_{-}\hat{E}_{+}\rangle=(c/V)\,\bar{n}_{\mathrm{th}}\hbar\omega_{\mathrm{d}}$. 

\begin{figure}
\includegraphics[width=1\columnwidth]{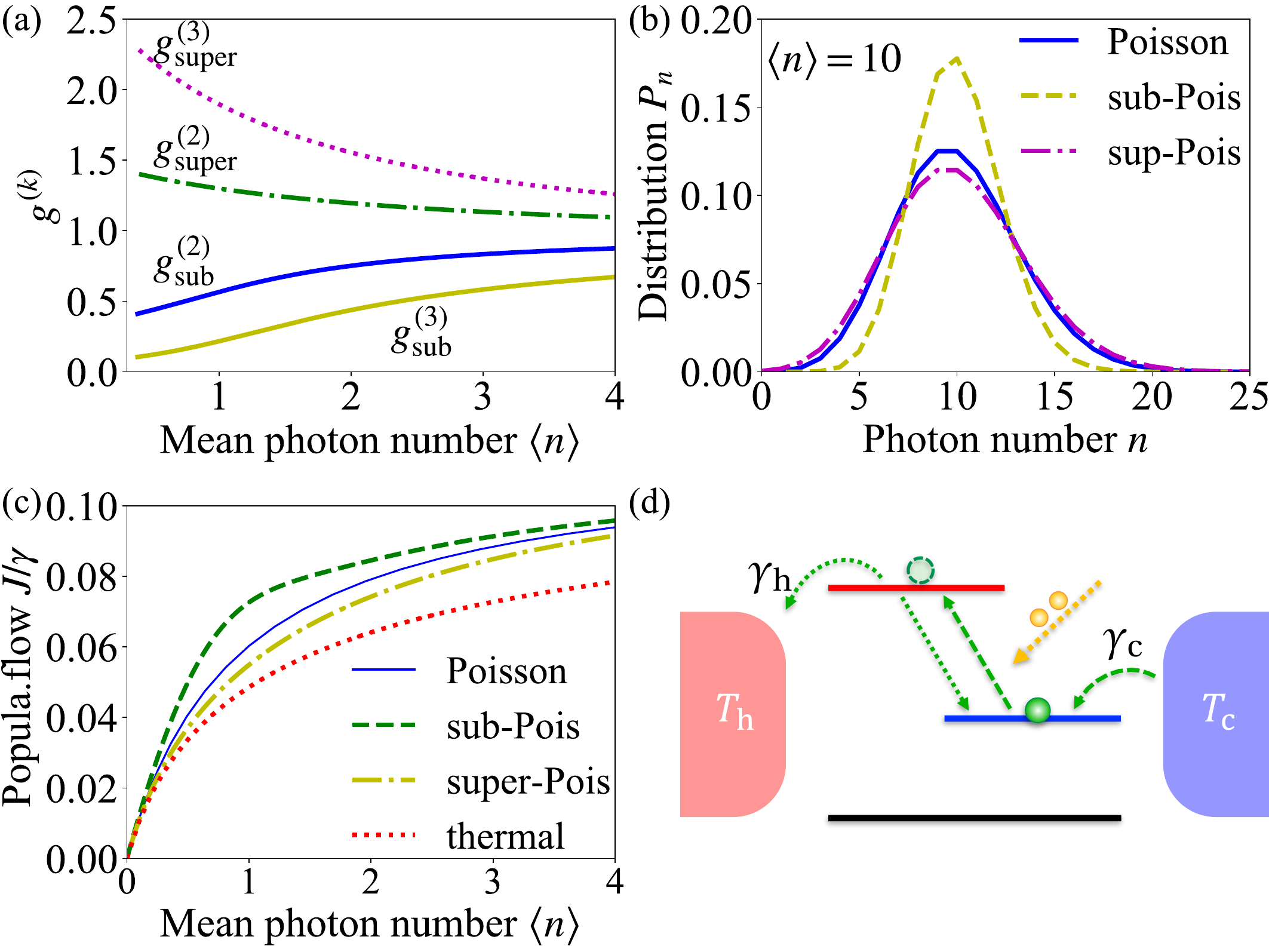}\caption{(a) The optical coherence $g^{(k)}$ for the photon distributions
(\ref{eq:Pn+-}). (b) Demonstration for the Poissonian, sub- and super-Poissonian
statistics (\ref{eq:Pn+-}) with the same mean photon number $\langle n\rangle=10$.
(c) The cooling power under different driving light obtained by Eq.
(\ref{eq:J-P}). Here we set $\xi_{0}/\gamma=1.1$ and $\bar{\mathsf{n}}_{\mathrm{c}}=0.5,\bar{\mathsf{n}}_{\mathrm{h}}=1$.
(d) Demonstration for the heat current blockage by a pair of bunching
photons. When two photons income together, the first photon would
excite the system up, but the second photon successively followed
would induce the stimulated emission, which draws down the energy
back before it flows to the hot bath, and that decreases the cooling
power.}
\label{fig-flows}
\end{figure}

It turns out the monochromatic ``thermal'' light is also driving
the system to work as a refrigerator rather than warming it up. But
comparing with the coherent driving {[}Eq.\,(\ref{eq:Q-hc}){]} under
the same light intensity, the driving light with thermal statistics
generates a smaller cooling power {[}see Fig.\,\ref{fig-flows}(c){]},
while the COP keeps the same as Eq.\,(\ref{eq:cofficient}).

A more attracting situation is when the driving light has nonclassical
photon statistics, e.g., the antibunching light. For nonclassical
light states, their P functions may contain negative parts, or could
be highly singular functions, thus cannot be regarded as legal probability
distributions \citep{orszag_quantum_2000,vogel_quantum_2006,scully_quantum_1997,agarwal_quantum_2012}.
In these cases, besides adopting the specific P function directly,
the above average (\ref{eq:J-P}) also can be calculated in the following
way: since the P function average is equal to the normal order expectation
by making the replacement $\alpha,\alpha^{*}\rightarrow\hat{a},\hat{a}^{\dagger}$,
$|\vec{E}_{\alpha}|^{2}\rightarrow\hat{E}_{-}\hat{E}_{+}$ \citep{glauber_coherent_1963,sudarshan_equivalence_1963},
the above population flow becomes \citep{widder_convolution_1954,agarwal_quantum_2012,yao_enhancing_2020}
\begin{align}
\overline{J} & =\big\langle:\frac{\mathscr{A}\hat{E}_{-}\hat{E}_{+}}{\mathscr{B}+\mathscr{C}\hat{E}_{-}\hat{E}_{+}}:\big\rangle\nonumber \\
 & =\big\langle:\frac{\mathscr{A}}{\mathscr{C}}-\frac{\mathscr{A}}{\mathscr{C}}\int_{0}^{\infty}ds\,e^{-s(1+\frac{\mathscr{C}}{\mathscr{B}}\hat{E}_{-}\hat{E}_{+})}:\big\rangle.\label{eq:normalOrder}
\end{align}
Here $\langle:\hat{J}(\hat{a}^{\dagger},\hat{a}):\rangle$ denotes
the normal order expectation. Thus, once the generation function $F(\tilde{s})\equiv\langle:\exp(-\tilde{s}\hat{a}^{\dagger}\hat{a}):\rangle$
is obtained for the driving light statistics, the heat flows of the
refrigerator can be calculated from the above integral.

Here we consider two examples of the photon statistics for the driving
light, i.e., 
\begin{equation}
P_{n}^{(-)}=\frac{1}{Z_{-}}\frac{\lambda^{n}}{(2n)!},\quad P_{n}^{(+)}=\frac{1}{Z_{+}}\frac{\lambda^{n}}{(n+2)!},\label{eq:Pn+-}
\end{equation}
with $Z_{\pm}(\lambda)$ as the normalization factors (see Appendix
\ref{sec:Normal-order}). By checking the mean photon number $\langle n\rangle$
and variance $\langle\delta n^{2}\rangle$, we can verify $P_{n}^{(+)}$
and $P_{n}^{(-)}$ are super- and sub-Poissonian distributions respectively
{[}see Fig.\,\ref{fig-flows}(a, b){]}.

The cooling currents generated by these two types of photon statistics
are shown in Fig.\,\ref{fig-flows}(c). Comparing with the coherent
light with the same intensity, the driving light with super(sub)-Poissonian
photon statistics produces a smaller (stronger) cooling power; with
the increase of the driving light intensity, the cooling powers generated
by the super- and sub-Poissonian lights both converge to the one generated
by the coherent light. In contrast, the cooling power generated by
the thermal light always keeps a finite difference lower than the
coherent light situation.

This result can be explained with the help of the following expansion
for the above normal order expectation (\ref{eq:normalOrder}), 
\begin{align}
\overline{J} & =\mathscr{A}\sum_{k=1}^{\infty}\frac{(-\mathscr{C})^{k-1}}{\mathscr{B}^{k}}\langle\hat{E}_{-}^{k}\hat{E}_{+}^{k}\rangle\nonumber \\
 & =\frac{\mathscr{A}}{2c\epsilon_{0}\mathscr{B}}I_{E}-\frac{\mathscr{AC}}{(2c\epsilon_{0}\mathscr{B})^{2}}I_{E}^{2}\cdot g^{(2)}+\dots\label{eq:J-g(2)}
\end{align}
Here $g^{(k)}\equiv\langle\hat{E}_{-}^{k}\hat{E}_{+}^{k}\rangle\big/\langle\hat{E}_{-}\hat{E}_{+}\rangle^{k}$
is the $k$-order coherence function of the driving light (zero time),
and $I_{E}\equiv2c\epsilon_{0}\langle\hat{E}_{-}\hat{E}_{+}\rangle$
is the light intensity. 

It turns out, when the light intensity is not too strong, the heat
flow induced by the driving light is proportional to the light intensity
$I_{E}$ in spite of the photon statistics {[}see the first term of
Eq.\,(\ref{eq:J-g(2)}){]}. With the increase of the light intensity
$I_{E}$, the high order coherence $g^{(k)}$ ($k\ge2$) of the driving
light also take effect to the heat flow. From the minus sign of the
second term in Eq.\,(\ref{eq:J-g(2)}), comparing with the situation
of coherent driving $g^{(2)}=1$, the driving lights with the sub-Poisson
($g^{(2)}<1$) and super-Poisson ($g^{(2)}>1$) statistics would generate
larger and smaller flows respectively. 

Moreover, it is worth noting that, for the two types of photon statistics
(\ref{eq:Pn+-}), they both give $g_{\mathrm{sub/super}}^{(2)}\rightarrow1$
when the light intensities grow large. Therefore, the cooling powers
they generate converge to the coherent light situation. In contrast,
the thermal light always gives $g_{\mathrm{th}}^{(2)}=2$ in spite
of the light intensity, thus the cooling power generated by the thermal
light always keeps a finite difference lower than the coherent light
situation. 

This effect can be understood by the demonstration in Fig.\,\ref{fig-flows}(d).
Generally, one incoming photon would excite the system up $|\mathsf{e}_{1}\rangle\rightarrow|\mathsf{\mathsf{e}}_{2}\rangle$,
and then generate an energy flow to the hot bath. But if a pair of
bunching photons come together, after the system is excited to $|\mathsf{\mathsf{e}}_{2}\rangle$
by the first photon, the second photon successively followed could
immediately induce the simulated emission and draw back the system
to $|\mathsf{\mathsf{e}}_{1}\rangle$, which prevents the energy flowing
to the hot bath. Such a blockage effect depends on the competition
between the releasing rate to the hot bath ($\sim\gamma_{\mathrm{h}}$)
and the stimulated emission rate due to the incoming photons ($\sim\xi_{0}$).

For a driving light with the Poissonian statistics, the photons income
to the system randomly, either in bunches or individually, while the
bunching photon pairs would decrease the cooling current flowing from
the cold bath to the hot one. Therefore, comparing with the coherent
light with the Poissonian statistics, the bunching (antibunching)
light, which has the super(sub)-Poissonian statistics, would contribute
more (less) blocking effect due to the stimulated emission, and thus
produces a smaller (stronger) cooling power. It should be emphasized
that here we focus on the situation that the linewidth of the driving
light is negligible comparing with the decay rates $\gamma_{\mathrm{h,c}}$,
thus the corrections from the finite bandwidth of the driving light
are omitted.

\section{The whole light field as a thermal bath}

As a comparison, here we consider another situation that the multimode
EM field as a whole is a heat bath \citep{skrzypczyk_smallest_2011,chen_quantum_2012,levy_quantum_2012,wang_four-level_2015},
staying in the thermal equilibrium state $\boldsymbol{\rho}_{\text{\textsc{e}}}\propto\exp[-\hat{H}_{\text{\textsc{e}}}/k_{\text{\textsc{b}}}T_{\text{\textsc{e}}}]$
($T_{\text{\textsc{e}}}$ is the temperature), and there is no other
driving light beam {[}see Fig. \ref{fig-3level}(b){]}.

In this case, the incoherent thermal lights are injecting to the system
from all different directions with different frequencies. The system
dynamics is now described by the following master equation \citep{breuer_theory_2002},
\begin{align}
\partial_{t}\tilde{\rho} & =\mathcal{L}_{\text{\textsc{e}}}'[\tilde{\rho}]+\mathcal{L}_{\mathrm{c}}[\tilde{\rho}]+\mathcal{L}_{\mathrm{h}}[\tilde{\rho}],\nonumber \\
\mathcal{L}_{\text{\textsc{e}}}'[\tilde{\rho}] & =\kappa\bar{\mathsf{n}}_{\text{\textsc{e}}}\big(\hat{\sigma}^{+}\tilde{\rho}\hat{\sigma}^{-}-\frac{1}{2}\{\tilde{\rho},\,\hat{\sigma}^{-}\hat{\sigma}^{+}\}_{+}\big)\nonumber \\
 & +\kappa(\bar{\mathsf{n}}_{\text{\textsc{e}}}+1)\big(\hat{\sigma}^{-}\tilde{\rho}\hat{\sigma}^{+}-\frac{1}{2}\{\tilde{\rho},\,\hat{\sigma}^{+}\hat{\sigma}^{-}\}_{+}\big),\label{eq:ME-T}
\end{align}
where $\bar{\mathsf{n}}_{\text{\textsc{e}}}:=[\exp(\hbar\Omega/k_{\text{\textsc{b}}}T_{\text{\textsc{e}}})-1]^{-1}$
is the mean thermal photon number, and $\mathcal{L}_{\mathrm{h,c}}[\tilde{\rho}]$
are the same as Eq.\,(\ref{eq:Lindblad}) indicating the dissipation
due to the coupling with the hot and cold baths.

Similarly as Eq.\,(\ref{eq:Q-hc}), the energy flows from system
to the hot, cold, EM baths are gives by the above master equation,
i.e., $\partial_{t}\langle\hat{H}_{\text{\textsc{s}}}\rangle=\mathcal{Q}_{\text{\textsc{e}}}'+\mathcal{Q}_{\mathrm{c}}'+\mathcal{Q}_{\mathrm{h}}'$.
In the steady state, they give 
\begin{align}
\mathcal{Q}_{\mathrm{h}}' & =-\hbar\omega_{\mathrm{h}}J',\quad\mathcal{Q}_{\mathrm{c}}'=\hbar\omega_{\mathrm{c}}J',\quad\mathcal{Q}_{\text{\textsc{e}}}'=\hbar\Omega\,J',\nonumber \\
J' & =\frac{\kappa\Big[\bar{\mathsf{n}}_{\text{\textsc{e}}}(\bar{\mathsf{n}}_{\mathrm{c}}-\bar{\mathsf{n}}_{\mathrm{h}})-\bar{\mathsf{n}}_{\mathrm{h}}(\bar{\mathsf{n}}_{\mathrm{c}}+1)\Big]}{\mathcal{N}+\frac{\kappa}{\gamma}\mathcal{M}},\label{eq:current-T}\\
\mathcal{M} & =\bar{\mathsf{n}}_{\text{\textsc{e}}}(3\bar{\mathsf{n}}_{\mathrm{h}}+3\bar{\mathsf{n}}_{\mathrm{c}}+2)+\bar{\mathsf{n}}_{\mathrm{h}}+2\bar{\mathsf{n}}_{\mathrm{c}}+1,\nonumber 
\end{align}
and $\mathcal{N}$ is the same as Eq.\,(\ref{eq:Gamma}). 

To make the system work as refrigerator, namely, the heat flows from
the cold bath to the hot one, the above heat flows require $J'>0$,
and that gives 
\begin{equation}
\bar{\mathsf{n}}_{\mathrm{c}}-\bar{\mathsf{n}}_{\mathrm{h}}\ge\frac{\bar{\mathsf{n}}_{\mathrm{h}}(\bar{\mathsf{n}}_{\mathrm{c}}+1)}{\bar{\mathsf{n}}_{\text{\textsc{e}}}}\quad\Leftrightarrow\quad\frac{\omega_{\mathrm{c}}}{T_{\mathrm{c}}}+\frac{\Omega}{T_{\text{\textsc{e}}}}\le\frac{\omega_{\mathrm{h}}}{T_{\mathrm{h}}}\label{eq:cooling-T}
\end{equation}
by substituting the Planck functions. When this condition is not satisfied,
the heat flows from the hot bath to the cold one, and the system is
not working as a refrigerator. Since $\omega_{\mathrm{h}}-\omega_{\mathrm{c}}\equiv\Omega$,
this cooling condition requires that the three temperatures must satisfy
$T_{\mathrm{c}}<T_{\mathrm{h}}\le T_{\text{\textsc{e}}}$. 

Clearly, the cooling condition here is different from the situation
that the refrigerator is driven by a monochromatic light beam, which
only requires $\bar{\mathsf{n}}_{\mathrm{c}}-\bar{\mathsf{n}}_{\mathrm{h}}\ge0$
and nonzero intensity for the driving light {[}Eq.\,(\ref{eq:Q-hc}){]}.
Unlike the above monochromatic driving case, here the incoherent thermal
lights are coming to the system from all different directions with
different frequencies. It turns out the cooling condition here requires
that the mean thermal photon number $\bar{\mathsf{n}}_{\text{\textsc{e}}}(\Omega,T_{\text{\textsc{e}}})$
in the EM bath must be larger than a certain threshold $\bar{\mathsf{n}}_{\text{\textsc{e}}}\ge\bar{\mathsf{n}}_{\mathrm{h}}(\bar{\mathsf{n}}_{\mathrm{c}}+1)/(\bar{\mathsf{n}}_{\mathrm{c}}-\bar{\mathsf{n}}_{\mathrm{h}})$.
And the COP of this refrigerator is 
\begin{equation}
e=\frac{|\mathcal{Q}_{\mathrm{c}}'|}{|\mathcal{Q}_{\mathrm{h}}'|-|\mathcal{Q}_{\mathrm{c}}'|}=\frac{\omega_{\mathrm{c}}}{\omega_{\mathrm{h}}-\omega_{\mathrm{c}}}\le\frac{T_{\mathrm{c}}-T_{\mathrm{h}}T_{\mathrm{c}}/T_{\text{\textsc{e}}}}{T_{\mathrm{h}}-T_{\mathrm{c}}},
\end{equation}
whose upper bound is smaller than the above monochromatic driving
case {[}Eq.\,(\ref{eq:cofficient}){]}. Such an upper bound also
has been obtained in some previous studies about the quantum absorption
refrigerators working between three thermal baths \citep{skrzypczyk_smallest_2011,chen_quantum_2012,levy_quantum_2012}.
Clearly, the working status of the refrigerator is significantly dependent
on the the specific quantum state of the EM field, not only on the
incoming light intensity.

\section{Discussion }

In the paper, we study a three-level quantum refrigerator which is
driven by a generic light state, even a nonclassical one. Since the
driving light input to the refrigerator could be a generic quantum
state, it is no longer enough to treat the driving light simply as
a planar wave, which is a quasi-classical description in literatures.
With the help of the P function representation, a generic driving
state can be regarded as the combination of many coherent states $|\alpha\rangle$
with $P(\alpha,\alpha^{*})$ as the ``quasi-probability'', while
the coherent input states could well return the quasi-classical driving
description. Therefore, the full system dynamics can be obtained as
the P function average of many evolution ``branches'', and each
evolution branch is obtained from the quasi-classical approach by
treating the driving light as a planar wave.

Based on this approach, it turns out all different input light states
give the same COP for this refrigerator, while the cooling power depend
not only on the light intensity but also the specific photon statistics
of the driving light. Comparing with the coherent light with the same
intensity, the driving light with super(sub)-Poissonian photon statistics
could raise a smaller (stronger) cooling power. We find that this
is because the bunching photons could block the cooling current generation
due to the the spontaneous emission they enhanced. This mechanism
could provide a more delicate control method via the high order coherence
of the input light. 

As a comparison, we also consider the situation that the multimode
EM field as a whole is in the thermal equilibrium state, and the incoherent
thermal lights with different frequencies are injecting to the system
from all different directions. It turns out the thermal photon number
in the EM bath must be larger than a certain threshold so as to make
the system work as a refrigerator. Therefore, the working status of
the refrigerator is significantly dependent on the the photon statistics
and frequency distribution of the EM field, not only on the incoming
light intensity.

\vspace{0.5em}

\emph{Acknowledgments }- S.-W. Li appreciates quite much for the helpful
discussion with L.-P. Yang in Northeast Normal University. This study
is supported by NSF of China (Grant No. 11905007).

\appendix
\begin{widetext}

\section{The system dynamics under generic driving light\label{sec:Master-equation-derivation}}

When the photon statistics of the driving light is taken into consideration,
it is no longer enough to treat the driving light only as a classical
planar wave, and the light state should be described by the fully
quantized EM field. In this situation, the classical planar wave corresponds
to a coherent state in the driving mode. Considering the whole EM
field is in a multimode coherent state $\hat{\boldsymbol{\rho}}_{\mathnormal{\textsc{b}}}^{\{\boldsymbol{\alpha}\}}:=\big|\{\boldsymbol{\alpha}\}\big\rangle\big\langle\{\boldsymbol{\alpha}\}\big|=\otimes_{\mathbf{k}\varsigma}|\alpha_{\mathbf{k}\varsigma}\rangle\langle\alpha_{\mathbf{k}\varsigma}|$,
the expectation of the electric field operator {[}Eq.\,(\ref{eq:E-alpha}){]}
gives 
\begin{equation}
\langle\hat{\mathbf{E}}(\mathbf{x},t)\rangle=\sum_{\mathbf{k}\varsigma}\hat{\mathrm{e}}_{\mathbf{k}\varsigma}\sqrt{\frac{\hbar\omega_{\mathbf{k}}}{2\epsilon_{0}V}}\,\big(i\,\alpha_{\mathbf{k}\varsigma}e^{i\mathbf{k}\cdot\mathbf{x}-i\omega_{\mathbf{k}}t}+\mathbf{h.c.}\big):=\vec{E}_{\boldsymbol{\alpha}}(\mathbf{x},t),
\end{equation}
 which just corresponds to a classical wave package composed of many
field modes, with the amplitude and phase of each mode-$(\mathbf{k}\varsigma)$
determined by $\alpha_{\mathbf{k}\varsigma}$. For an idealistic monochromatic
driving light, only the driving mode is in the coherent state $|\alpha\rangle$
while all the other field modes are in the vacuum states, then the
electric field gives $\vec{E}_{\boldsymbol{\alpha}}(\mathbf{x},t)\rightarrow\vec{E}_{\mathrm{d}}\sin(\mathbf{k}_{\mathrm{d}}\cdot\mathbf{x}-\omega_{\mathrm{d}}t-\phi_{\alpha})$,
with the amplitude $\vec{E}_{\mathrm{d}}\equiv\hat{\mathrm{e}}_{\mathrm{d}}\,|\alpha|\sqrt{2\hbar\omega_{\mathrm{d}}/\epsilon_{0}V}$,
and $\phi_{\alpha}=\arg\alpha$.

To study the interaction with such an EM field in a multimode coherent
state, we start from the general interaction between a two-level system
and the fully quantized EM field, that is (interaction picture),
\begin{equation}
\tilde{H}_{\text{\textsc{sb}}}=-\tilde{\boldsymbol{d}}(t)\cdot\hat{\mathbf{E}}(\mathbf{x}_{0},t)=-\tilde{\boldsymbol{d}}(t)\cdot\sum_{\mathbf{k}\varsigma}\hat{\mathrm{e}}_{\mathbf{k}\varsigma}\sqrt{\frac{\hbar\omega_{\mathbf{k}}}{2\epsilon_{0}V}}\big(i\,\hat{a}_{\mathbf{k}\varsigma}\,e^{i\mathbf{k}\cdot\mathbf{x}_{0}-i\omega_{\mathbf{k}}t}+\mathbf{h.c.}\big),
\end{equation}
with $\mathbf{x}_{0}$ as the position of the two-level system (hereafter
we set $\mathbf{x}_{0}=0$). The field operators can be divided as
the summation of their mean values and quantum fluctuations, i.e.,
$\hat{a}_{\mathbf{k}\varsigma}\equiv\langle\hat{a}_{\mathbf{k}\varsigma}\rangle+\delta\hat{a}_{\mathbf{k}\varsigma}=\alpha_{\mathbf{k}\varsigma}+\delta\hat{a}_{\mathbf{k}\varsigma}$,
and $\hat{\mathbf{E}}\equiv\vec{E}_{\boldsymbol{\alpha}}+\delta\hat{\mathbf{E}}$,
then the above interaction also can be divided into two parts $\tilde{H}_{\text{\textsc{sb}}}=\tilde{V}_{\boldsymbol{\alpha}}(t)+\tilde{H}_{\text{\textsc{sb}}}^{(0)}$,
where 
\begin{align}
\tilde{V}_{\boldsymbol{\alpha}}(t) & =-\tilde{\boldsymbol{d}}(t)\cdot\vec{E}_{\boldsymbol{\alpha}}(\mathbf{x}_{0},t),\nonumber \\
\tilde{H}_{\text{\textsc{sb}}}^{(0)} & =-\tilde{\boldsymbol{d}}(t)\cdot\sum_{\mathbf{k}\varsigma}\hat{\mathrm{e}}_{\mathbf{k}\varsigma}\sqrt{\frac{\hbar\omega_{\mathbf{k}}}{2\epsilon_{0}V}}\big(i\,\delta\hat{a}_{\mathbf{k}\varsigma}\,e^{-i\omega_{\mathbf{k}}t}+\mathbf{h.c.}\big).
\end{align}
Notice that $\tilde{V}_{\boldsymbol{\alpha}}(t)$ just indicates the
interaction between the electric dipole and a planar wave $\vec{E}_{\boldsymbol{\alpha}}(\mathbf{x}_{0},t)$.

Further, the dynamics of the system can be obtained by taking the
integral iteration of the von Neumann equation (interaction picture),
that is, 
\begin{align}
\partial_{t}\tilde{\boldsymbol{\rho}}_{\text{\textsc{sb}}}(t) & =\frac{i}{\hbar}[\tilde{\boldsymbol{\rho}}_{\text{\textsc{sb}}}(t),\,\tilde{V}_{\boldsymbol{\alpha}}(t)]+\frac{i}{\hbar}[\tilde{\boldsymbol{\rho}}_{\text{\textsc{sb}}}(t),\,\tilde{H}_{\text{\textsc{sb}}}^{(0)}(t)]\nonumber \\
 & =\frac{i}{\hbar}[\tilde{\boldsymbol{\rho}}_{\text{\textsc{sb}}}(t),\,\tilde{V}_{\boldsymbol{\alpha}}(t)]+\frac{i}{\hbar}[\tilde{\boldsymbol{\rho}}_{\text{\textsc{sb}}}(0),\,\tilde{H}_{\text{\textsc{sb}}}^{(0)}(t)]-\frac{1}{\hbar^{2}}\int_{0}^{t}ds\,\Big[[\tilde{\boldsymbol{\rho}}_{\text{\textsc{sb}}}(s),\,\tilde{V}_{\boldsymbol{\alpha}}(s)+\tilde{H}_{\text{\textsc{sb}}}^{(0)}(s)],\,\tilde{H}_{\text{\textsc{sb}}}^{(0)}(t)\Big].
\end{align}
Here $\tilde{H}_{\text{\textsc{sb}}}^{(0)}$ only contains the contribution
of pure quantum fluctuation $\delta\hat{a}_{\mathbf{k}\varsigma}$,
which gives the same result as the situation dealing with the spontaneous
emission in the vacuum field. Thus, applying the Born-Markov-RWA,
the master equation for the system dynamics is obtained as \citep{breuer_theory_2002,yao_enhancing_2020}
\begin{equation}
\partial_{t}\tilde{\mathbf{\rho}}_{\text{\textsc{s}}}^{\{\boldsymbol{\alpha}\}}=\frac{i}{\hbar}[\tilde{\mathbf{\rho}}_{\text{\textsc{s}}}^{\{\boldsymbol{\alpha}\}},\tilde{V}_{\boldsymbol{\alpha}}(t)]+\kappa\big(\hat{\sigma}^{-}\tilde{\mathbf{\rho}}_{\text{\textsc{s}}}^{\{\boldsymbol{\alpha}\}}\hat{\sigma}^{+}-\frac{1}{2}\hat{\sigma}^{+}\hat{\sigma}^{-}\tilde{\mathbf{\rho}}_{\text{\textsc{s}}}^{\{\boldsymbol{\alpha}\}}-\frac{1}{2}\tilde{\mathbf{\rho}}_{\text{\textsc{s}}}^{\{\boldsymbol{\alpha}\}}\hat{\sigma}^{+}\hat{\sigma}^{-}\big).\label{eq:ME-alpha}
\end{equation}
The superscript $\{\boldsymbol{\alpha}\}$ indicates the initial state
of the EM field is the specific multimode coherent state $\hat{\boldsymbol{\rho}}_{\mathnormal{\textsc{b}}}^{\{\boldsymbol{\alpha}\}}$.
This is just the master equation dealing with quasi-classical driving
widely adopted in literature.

The above derivations also indicate that if the initial state of the
EM field is not a coherent state but a general one, the above master
equation (\ref{eq:ME-alpha}) for quasi-classical driving is not sufficient
enough to enclose the photon statistics of the driving light. In this
situation, generally, the field state always can be represented as
the following multimode P function,
\begin{equation}
\hat{\boldsymbol{\rho}}_{\mathnormal{\textsc{b}}}(0)=\int d\{\boldsymbol{\alpha},\boldsymbol{\alpha}^{*}\}\,\boldsymbol{P}\big(\{\boldsymbol{\alpha},\boldsymbol{\alpha}^{*}\}\big)\,\big|\{\boldsymbol{\alpha}\}\big\rangle\big\langle\{\boldsymbol{\alpha}\}\big|=\int d\{\boldsymbol{\alpha},\boldsymbol{\alpha}^{*}\}\,\boldsymbol{P}\big(\{\boldsymbol{\alpha},\boldsymbol{\alpha}^{*}\}\big)\,\hat{\boldsymbol{\rho}}_{\mathnormal{\textsc{b}}}^{\{\boldsymbol{\alpha}\}}.
\end{equation}
Formally, the density state $\hat{\boldsymbol{\rho}}_{\mathnormal{\textsc{b}}}(0)$
could be regarded as the combination of different coherent states
$\hat{\boldsymbol{\rho}}_{\mathnormal{\textsc{b}}}^{\{\boldsymbol{\alpha}\}}$
with $\boldsymbol{P}\big(\{\boldsymbol{\alpha},\boldsymbol{\alpha}^{*}\}\big)$
as the quasi-probability, and each $\hat{\boldsymbol{\rho}}_{\mathnormal{\textsc{b}}}^{\{\boldsymbol{\alpha}\}}$
corresponds to a classical wave package $\vec{E}_{\boldsymbol{\alpha}}(\mathbf{x},t)$.
Then the system dynamics $\hat{\rho}_{\text{\textsc{s}}}(t)=\mathrm{tr}\big[\hat{\boldsymbol{\rho}}_{\text{\textsc{sb}}}(t)\big]$
can be written as 
\begin{align}
\hat{\rho}_{\text{\textsc{s}}}(t) & =\mathrm{tr}_{\text{\textsc{b}}}\big[\mathcal{U}\,\hat{\rho}_{\text{\textsc{s}}}(0)\otimes\hat{\boldsymbol{\rho}}_{\text{\textsc{b}}}(0)\,\mathcal{U}^{\dagger}\big]=\int d\{\boldsymbol{\alpha},\boldsymbol{\alpha}^{*}\}\,\boldsymbol{P}\big(\{\boldsymbol{\alpha},\boldsymbol{\alpha}^{*}\}\big)\,\mathrm{tr}_{\text{\textsc{b}}}\big[\mathcal{U}\,\hat{\rho}_{\text{\textsc{s}}}(0)\otimes\hat{\boldsymbol{\rho}}_{\mathnormal{\textsc{b}}}^{\{\boldsymbol{\alpha}\}}\,\mathcal{U}^{\dagger}\big]\nonumber \\
 & =\int d\{\boldsymbol{\alpha},\boldsymbol{\alpha}^{*}\}\,\boldsymbol{P}\big(\{\boldsymbol{\alpha},\boldsymbol{\alpha}^{*}\}\big)\,\hat{\rho}_{\text{\textsc{s}}}^{\{\boldsymbol{\alpha}\}}(t).
\end{align}
Here $\hat{\rho}_{\text{\textsc{s}}}^{\{\boldsymbol{\alpha}\}}(t)\equiv\mathrm{tr}_{\text{\textsc{b}}}\big[\mathcal{U}\,\hat{\rho}_{\text{\textsc{s}}}(0)\otimes\hat{\boldsymbol{\rho}}_{\mathnormal{\textsc{b}}}^{\{\boldsymbol{\alpha}\}}\,\mathcal{U}^{\dagger}\big]$
indicates the system evolution when the initial state of the EM field
is $\hat{\boldsymbol{\rho}}_{\mathnormal{\textsc{b}}}^{\{\boldsymbol{\alpha}\}}=\big|\{\boldsymbol{\alpha}\}\big\rangle\big\langle\{\boldsymbol{\alpha}\}\big|$,
which just can be given by the above master equation (\ref{eq:ME-alpha}).
Correspondingly, the observable expectations of the system also can
be obtained as a P function average, that is, 
\begin{equation}
\langle\hat{o}_{\text{\textsc{s}}}(t)\rangle=\int d\{\boldsymbol{\alpha},\boldsymbol{\alpha}^{*}\}\,\boldsymbol{P}\big(\{\boldsymbol{\alpha},\boldsymbol{\alpha}^{*}\}\big)\,\langle\hat{o}_{\text{\textsc{s}}}^{(\boldsymbol{\alpha})}(t)\rangle,
\end{equation}
 where $\langle\hat{o}_{\text{\textsc{s}}}^{(\boldsymbol{\alpha})}(t)\rangle:=\mathrm{tr}_{\text{\textsc{s}}}[\hat{o}_{\text{\textsc{s}}}\cdot\hat{\rho}_{\text{\textsc{s}}}^{\{\boldsymbol{\alpha}\}}(t)]$
can be obtained the master equation (\ref{eq:ME-alpha}). Namely,
the fully system dynamics $\langle\hat{o}_{\text{\textsc{s}}}(t)\rangle$
could be regarded as the probabilistic combination of many evolution
``branches'' $\langle\hat{o}_{\text{\textsc{s}}}^{(\boldsymbol{\alpha})}(t)\rangle$
with $\boldsymbol{P}\big(\{\boldsymbol{\alpha},\boldsymbol{\alpha}^{*}\}\big)$
as the quasi-probability, and each ``branch'' can be obtained by
the quasi-classical driving approach.

\section{The energy flow under coherent driving \label{sec:The-power-of}}

Here we show the results for the steady state solution of the master
equation (\ref{eq:ME}), which describes the system driven by the
coherent light (modeled by a planar wave). Denoting $\Delta\equiv\Omega-\omega_{\mathrm{d}}$
as the detuning between the driving light and the transition frequency
$\hbar\Omega\equiv\text{\textsc{e}}_{2}-\text{\textsc{e}}_{1}$, under
the interaction picture defined by $\hat{H}_{\Delta}:=\hat{H}_{S}-\Delta|\mathsf{e}_{2}\rangle\langle\mathsf{e}_{2}|$,
the expectations of $\hat{\text{\textsc{n}}}_{1(2)}\equiv|\mathsf{e}_{1(2)}\rangle\langle\mathsf{e}_{1(2)}|$,
$\hat{\text{\textsc{n}}}_{\mathsf{g}}\equiv|\mathsf{g}\rangle\langle\mathsf{g}|$
and $\hat{\tau}_{\text{\textsc{e}}}^{+}\equiv(\hat{\tau}_{\text{\textsc{e}}}^{-})^{\dagger}\equiv|\mathsf{e}_{2}\rangle\langle\mathsf{e}_{1}|$)
form a closed set of time-independent equations, which read (setting
$\gamma_{\mathrm{h}}=\gamma_{\mathrm{c}}\equiv\gamma$)
\begin{alignat}{1}
\partial_{t}\langle\hat{\text{\textsc{n}}}_{1}\rangle & =-[\mathcal{E}\langle\hat{\tau}_{\textrm{\textsc{e}}}^{+}\rangle+\mathcal{E}^{*}\langle\hat{\tau}_{\textrm{\textsc{e}}}^{-}\rangle]-\gamma(\bar{\mathsf{n}}_{\mathrm{c}}+1)\langle\hat{\text{\textsc{n}}}_{1}\rangle+\kappa\langle\hat{\text{\textsc{n}}}_{2}\rangle+\gamma\bar{\mathsf{n}}_{\mathrm{c}}\langle\hat{\text{\textsc{n}}}_{\mathsf{g}}\rangle,\nonumber \\
\partial_{t}\langle\hat{\text{\textsc{n}}}_{2}\rangle & =[\mathcal{E}\langle\hat{\tau}_{\textrm{\textsc{e}}}^{+}\rangle+\mathcal{E}^{*}\langle\hat{\tau}_{\textrm{\textsc{e}}}^{-}\rangle]-\gamma(\bar{\mathsf{n}}_{\mathrm{h}}+1)\langle\hat{\text{\textsc{n}}}_{2}\rangle-\kappa\langle\hat{\text{\textsc{n}}}_{2}\rangle+\gamma\bar{\mathsf{n}}_{\mathrm{h}}\langle\hat{\text{\textsc{n}}}_{\mathsf{g}}\rangle,\\
\partial_{t}\langle\hat{\tau}_{\textrm{\textsc{e}}}^{+}\rangle & =i\Delta\langle\hat{\tau}_{\textrm{\textsc{e}}}^{+}\rangle+\mathcal{E}^{*}[\langle\hat{\text{\textsc{n}}}_{1}\rangle-\langle\hat{\text{\textsc{n}}}_{2}\rangle]-\frac{1}{2}[\gamma(\bar{\mathsf{n}}_{\mathrm{c}}+1)+\gamma(\bar{\mathsf{n}}_{\mathrm{h}}+1)+\kappa]\langle\hat{\tau}_{\textrm{\textsc{e}}}^{+}\rangle,\nonumber 
\end{alignat}
and $\langle\hat{\text{\textsc{n}}}_{1}\rangle+\langle\hat{\text{\textsc{n}}}_{2}\rangle+\langle\hat{\text{\textsc{n}}}_{\mathsf{g}}\rangle\equiv1$.
In the steady state $t\rightarrow\infty$, the system becomes stationary
and the above time derivatives give zero. Here we consider the situation
that the spontaneous emission rate $\kappa$ is negligible comparing
with the coupling strengths with the two heat baths, i.e., $\gamma\gg\kappa\simeq0$,
and then the steady state solution gives 
\begin{align}
\langle\hat{\text{\textsc{n}}}_{1}\rangle & =\frac{4\triangle^{2}\gamma^{2}\bar{\mathsf{n}}_{\mathrm{c}}(1+\bar{\mathsf{n}}_{\mathrm{h}})+4|\mathcal{E}|^{2}\gamma(\bar{\mathsf{n}}_{\mathrm{c}}+\bar{\mathsf{n}}_{\mathrm{h}})\Gamma_{1}+\gamma^{2}\bar{\mathsf{n}}_{\mathrm{c}}(1+\bar{\mathsf{n}}_{\mathrm{h}})\Gamma_{1}^{2}}{\Phi},\nonumber \\
\langle\hat{\text{\textsc{n}}}_{2}\rangle & =\frac{4\triangle^{2}\gamma^{2}\bar{\mathsf{n}}_{\mathrm{h}}(1+\bar{\mathsf{n}}_{\mathrm{c}})+4|\mathcal{E}|^{2}\gamma(\bar{\mathsf{n}}_{\mathrm{c}}+\bar{\mathsf{n}}_{\mathrm{h}})\Gamma_{1}+\gamma^{2}\bar{\mathsf{n}}_{\mathrm{h}}(1+\bar{\mathsf{n}}_{\mathrm{c}})\Gamma_{1}^{2}}{\Phi},\label{eq:steady-results}\\
\langle\hat{\tau}_{\textrm{\textrm{\textsc{e}}}}^{+}\rangle & =\big[\langle\hat{\tau}_{\textrm{\textrm{\textsc{e}}}}^{-}\rangle\big]^{*}=\frac{4i\triangle\,\mathcal{E}^{*}\gamma^{2}(\bar{\mathsf{n}}_{\mathrm{c}}-\bar{\mathsf{n}}_{\mathrm{h}})+2\mathcal{E}^{*}\gamma^{2}(\bar{\mathsf{n}}_{\mathrm{c}}-\bar{\mathsf{n}}_{\mathrm{h}})\Gamma_{1}}{\Phi},\nonumber 
\end{align}
where $\Phi=4\triangle^{2}\gamma^{2}\mathcal{N}+\Gamma_{1}(\Gamma_{1}\gamma^{2}\mathcal{N}+4|\mathcal{E}|^{2}\Gamma_{2})$,
and 
\begin{align}
\Gamma_{1} & =\gamma(2+\bar{\mathsf{n}}_{\mathrm{c}}+\bar{\mathsf{n}}_{\mathrm{h}}),\qquad\Gamma_{2}=\gamma(2+3\bar{\mathsf{n}}_{\mathrm{c}}+3\bar{\mathsf{n}}_{\mathrm{h}}),\nonumber \\
\mathcal{N} & =1+2\bar{\mathsf{n}}_{\mathrm{h}}+2\bar{\mathsf{n}}_{\mathrm{c}}+3\bar{\mathsf{n}}_{\mathrm{c}}\bar{\mathsf{n}}_{\mathrm{h}}.
\end{align}

Based on the master equation, the heat flows defined in Eq.\,(\ref{eq:E-Q})
give 
\begin{align}
\mathcal{Q}_{\mathrm{h(c)}} & =\hbar\omega_{\mathrm{h(c)}}\cdot\gamma_{\mathrm{h(c)}}\big[\bar{\mathsf{n}}_{\mathrm{h(c)}}\langle\hat{\text{\textsc{n}}}_{\mathsf{g}}\rangle-(\bar{\mathsf{n}}_{\mathrm{h(c)}}+1)\langle\hat{\text{\textsc{n}}}_{2(1)}\rangle\big],\nonumber \\
\mathcal{Q}_{\text{\textsc{e}}} & =\hbar\Omega\cdot\big[\mathcal{E}\langle\hat{\tau}_{\textrm{\textsc{e}}}^{+}\rangle+\mathcal{E}^{*}\langle\hat{\tau}_{\textrm{\textsc{e}}}^{-}\rangle\big]-\hbar\Omega\cdot\kappa\langle\hat{\text{\textsc{n}}}_{2}\rangle.
\end{align}
By substituting the steady state solutions (\ref{eq:steady-results})
into the above flows, they give the steady heat flows as Eq.\,(\ref{eq:Q-hc})
in the main text.

\section{The normal order expectation for the heat flow generated by generic
photon statistics \label{sec:Normal-order}}

Here we show how to calculate the normal order expectation for the
population flow in Eq.\,(\ref{eq:normalOrder}). Notice that the
population flow has been turned into the integral of the normal order
expectation $\langle:e^{-\tilde{s}\hat{a}^{\dagger}\hat{a}}:\rangle\equiv F(\tilde{s})$,
which is the critical part to be calculated. Here we consider the
density state of the monochromatic driving light is diagonal in the
Fock basis, i.e., $\rho=\sum P_{n}|n\rangle\langle n|$ with $P_{n}$
as the photon number distribution, thus the characteristic function
$F(\tilde{s})$ gives 
\begin{align}
F(\tilde{s}) & =\sum_{k=0}^{\infty}\frac{(-\tilde{s})^{k}}{k!}\langle(\hat{a}^{\dagger})^{k}\hat{a}^{k}\rangle=\sum_{k=0}^{\infty}\sum_{m,n=0}^{\infty}\rho_{mn}\frac{(-\tilde{s})^{k}}{k!}\langle n|(\hat{a}^{\dagger})^{k}\hat{a}^{k}|m\rangle\nonumber \\
 & =\sum_{n=0}^{\infty}\sum_{k=0}^{n}\rho_{nn}\cdot\frac{(-\tilde{s})^{k}n!}{k!(n-k)!}=\sum_{n}P_{n}(1-\tilde{s})^{n}.
\end{align}
For the examples of the Poisson, sub-Poisson and super-Poisson statistics
discussed in the main text, they are 
\begin{alignat}{2}
\text{Poisson: } & P_{n}=\frac{1}{e^{\lambda}}\frac{\lambda^{n}}{n!}, & \quad\Rightarrow\; & F(\tilde{s})=e^{-\tilde{s}\lambda},\nonumber \\
\text{sub-Poisson: } & P_{n}^{(-)}=\frac{1}{Z_{-}(\lambda)}\frac{\lambda^{n}}{(2n)!}, & \quad\Rightarrow\; & F(\tilde{s})=\frac{Z_{-}\big(\,\lambda(1-\tilde{s})\,\big)}{Z_{-}(\lambda)},\quad Z_{-}(\lambda)\equiv\cosh\sqrt{\lambda},\\
\text{super-Poisson: } & P_{n}^{(+)}=\frac{1}{Z_{+}(\lambda)}\frac{\lambda^{n}}{(n+2)!} & \quad\Rightarrow\; & F(\tilde{s})=\frac{Z_{+}\big(\,\lambda(1-\tilde{s})\,\big)}{Z_{+}(\lambda)},\quad Z_{+}(\lambda)\equiv\frac{e^{\lambda}-\lambda-1}{\lambda^{2}}.\nonumber 
\end{alignat}
Then the population flow (\ref{eq:normalOrder}) can be further calculated
from the integral numerically, i.e.,
\begin{align}
\overline{J} & =\big\langle:\frac{\mathscr{A}\hat{E}_{-}\hat{E}_{+}}{\mathscr{B}+\mathscr{C}\hat{E}_{-}\hat{E}_{+}}:\big\rangle=\big\langle:\frac{\mathscr{A}}{\mathscr{C}}-\frac{\mathscr{A}}{\mathscr{C}}\int_{0}^{\infty}ds\,e^{-s(1+\frac{\mathscr{C}}{\mathscr{B}}\hat{E}_{-}\hat{E}_{+})}:\big\rangle\nonumber \\
 & =\frac{\mathscr{A}}{\mathscr{C}}-\frac{\mathscr{A}}{\mathscr{C}}\int_{0}^{\infty}ds\,e^{-s}F\Big(\frac{\mathscr{C}}{\mathscr{B}}\frac{\hbar\omega_{\mathrm{d}}}{2\epsilon_{0}V}s\Big).
\end{align}

\section{The energy flow generated by the multimode thermal field }

Here we show the results for the energy flow when the whole multimode
EM field is in the thermal equilibrium state $\boldsymbol{\rho}_{\text{\textsc{e}}}\propto\exp[-\hat{H}_{\text{\textsc{e}}}/k_{\text{\textsc{b}}}T_{\text{\textsc{e}}}]$
with $T_{\text{\textsc{e}}}$ as the temperature. In this case the
incoherent thermal lights with different frequencies are injecting
to the system from all different directions, and there is no other
driving light beam. The system dynamics is described by the master
equation (\ref{eq:ME-T}), and that gives 
\begin{align}
\partial_{t}\langle\hat{\text{\textsc{n}}}_{1}\rangle= & \kappa\big[(\bar{\mathsf{n}}_{\text{\textsc{e}}}+1)\langle\hat{\text{\textsc{n}}}_{2}\rangle-\bar{\mathsf{n}}_{\text{\textsc{e}}}\langle\hat{\text{\textsc{n}}}_{1}\rangle\big]-\gamma\big[(\bar{\mathsf{n}}_{\mathrm{c}}+1)\langle\hat{\text{\textsc{n}}}_{1}\rangle-\bar{\mathsf{n}}_{\mathrm{c}}\langle\hat{\text{\textsc{n}}}_{\mathsf{g}}\rangle\big],\nonumber \\
\partial_{t}\langle\hat{\text{\textsc{n}}}_{2}\rangle= & -\kappa\big[(\bar{\mathsf{n}}_{\text{\textsc{e}}}+1)\langle\hat{\text{\textsc{n}}}_{2}\rangle-\bar{\mathsf{n}}_{\text{\textsc{e}}}\langle\hat{\text{\textsc{n}}}_{1}\rangle\big]-\gamma\big[(\bar{\mathsf{n}}_{\mathrm{h}}+1)\langle\hat{\text{\textsc{n}}}_{2}\rangle-\bar{\mathsf{n}}_{\mathrm{h}}\langle\hat{\text{\textsc{n}}}_{\mathsf{g}}\rangle\big].
\end{align}
In the steady state $t\rightarrow\infty$, $\partial_{t}\langle\hat{\text{\textsc{n}}}_{1,2}\rangle=0$
and $\langle\hat{\text{\textsc{n}}}_{1}\rangle+\langle\hat{\text{\textsc{n}}}_{2}\rangle+\langle\hat{\text{\textsc{n}}}_{\mathsf{g}}\rangle\equiv1$
give the solution as 
\begin{align}
\langle\hat{\text{\textsc{n}}}_{1}\rangle & =\frac{\bar{\mathsf{n}}_{\mathrm{c}}(\bar{\mathsf{n}}_{\mathrm{h}}+1)+\frac{\kappa}{\gamma}(\bar{\mathsf{n}}_{\text{\textsc{e}}}+1)(\bar{\mathsf{n}}_{\mathrm{h}}+\bar{\mathsf{n}}_{\mathrm{c}})}{\mathcal{N}+\frac{\kappa}{\gamma}\mathcal{M}},\nonumber \\
\langle\hat{\text{\textsc{n}}}_{2}\rangle & =\frac{\bar{\mathsf{n}}_{\mathrm{h}}(\bar{\mathsf{n}}_{\mathrm{c}}+1)+\frac{\kappa}{\gamma}(\bar{\mathsf{n}}_{\text{\textsc{e}}}+1)(\bar{\mathsf{n}}_{\mathrm{h}}+\bar{\mathsf{n}}_{\mathrm{c}})}{\mathcal{N}+\frac{\kappa}{\gamma}\mathcal{M}},\quad\mathcal{M}\equiv\bar{\mathsf{n}}_{\text{\textsc{e}}}(3\bar{\mathsf{n}}_{\mathrm{h}}+3\bar{\mathsf{n}}_{\mathrm{c}}+2)+\bar{\mathsf{n}}_{\mathrm{h}}+2\bar{\mathsf{n}}_{\mathrm{c}}+1,
\end{align}
and $\mathcal{N}$ is the same as Eq.\,(\ref{eq:Gamma}) in the coherent
driving case. With the help of the master equation (\ref{eq:ME-T}),
the energy flows from the three reservoirs to the system is defined
from the conservation relation $\partial_{t}\langle\hat{H}_{\text{\textsc{s}}}\rangle=\mathcal{Q}_{\text{\textsc{e}}}'+\mathcal{Q}_{\mathrm{c}}'+\mathcal{Q}_{\mathrm{h}}'$.
In the steady state, they give 
\begin{align}
\mathcal{Q}_{\mathrm{h}}' & =-\hbar\omega_{\mathrm{h}}J',\quad\mathcal{Q}_{\mathrm{c}}'=\hbar\omega_{\mathrm{c}}J',\quad\mathcal{Q}_{\text{\textsc{e}}}'=\hbar\Omega\,J',\nonumber \\
J' & =\frac{\kappa\Big[\bar{\mathsf{n}}_{\text{\textsc{e}}}(\bar{\mathsf{n}}_{\mathrm{c}}-\bar{\mathsf{n}}_{\mathrm{h}})-\bar{\mathsf{n}}_{\mathrm{h}}(\bar{\mathsf{n}}_{\mathrm{c}}+1)\Big]}{\mathcal{N}+\frac{\kappa}{\gamma}\mathcal{M}},
\end{align}
When $J'>0$, the heat current flows from the cold bath to the hot
one, and thus the system works as a refrigerator, and that requires
the cooling condition (\ref{eq:cooling-T}) in the main text.

\end{widetext}

\end{document}